\documentclass[sigconf,authorversion,nonacm]{acmart}

\AtBeginDocument{%
  }

\newcommand{\etal}{\hbox{et al.}\xspace}
\newcommand{\eg}{\hbox{e.g.}\xspace}

\newcommand{\junjie}[1]{\textcolor{orange}{#1}}
\newcommand{\mao}[1]{\textcolor{blue}{#1}}

\newcommand{\chen}[1]{\textcolor{red}{#1}}

\newcommand{\llamaversion}{llama3-70b-8192\xspace}
\newcommand{\gptversion}{gpt-4o\xspace}
\newcommand{\llmapp}{LLMapp\xspace}
\newcommand{\llmapps}{LLMapps\xspace}

\definecolor{TUMBlue}{HTML}{0065BD}
\definecolor{TUMSecondaryBlue}{HTML}{005293}
\definecolor{TUMSecondaryBlue2}{HTML}{003359}
\definecolor{TUMBlack}{HTML}{000000}
\definecolor{TUMWhite}{HTML}{FFFFFF}
\definecolor{TUMDarkGray}{HTML}{333333}
\definecolor{TUMGray}{HTML}{808080}
\definecolor{TUMLightGray}{HTML}{CCCCC6}
\definecolor{TUMAccentGray}{HTML}{DAD7CB}
\definecolor{TUMAccentOrange}{HTML}{E37222}
\definecolor{TUMAccentGreen}{HTML}{A2AD00}
\definecolor{TUMAccentLightBlue}{HTML}{98C6EA}
\definecolor{TUMAccentBlue}{HTML}{64A0C8}

\usepackage{xcolor}
\usepackage{cleveref}
\usepackage{multirow}
\usepackage{subfig}
\usepackage{graphicx}
\usepackage{listings}
\usepackage[most]{tcolorbox}
\usepackage{ulem}
\usepackage{threeparttable}
\usepackage{colortbl}
\usepackage{enumitem}
\usepackage{xspace}

\setcopyright{acmlicensed}
\copyrightyear{2025}
\acmYear{2025}
\acmDOI{XXXXXXX.XXXXXXX}
\acmConference[Conference acronym 'XX]{Make sure to enter the correct
  conference title from your rights confirmation email}{June 03--05,
  2018}{Woodstock, NY}
\acmBooktitle{Companion Proceedings of the 33rd ACM Symposium on the Foundations of Software Engineering (FSE '25), June 23--27, 2025, Trondheim, Norway}
\acmISBN{978-1-4503-XXXX-X/18/06}
\received{2025-XX-XX}
\received[accepted]{2025-XX-XX}

\begin{document}

\title{From Prompts to Templates: A Systematic Prompt Template Analysis for Real-world \llmapps}


\author{Yuetian Mao}
\affiliation{%
  \institution{Technical University of Munich}
  \country{Germany}}
\email{yuetian.mao@tum.de}

\author{Junjie He}
\affiliation{%
  \institution{Technical University of Munich}
  \country{Germany}}
\email{junjie.he@tum.de}

\author{Chunyang Chen}
\authornote{Chunyang Chen is the corresponding author.}
\affiliation{%
  \institution{Technical University of Munich}
  \country{Germany}}
\email{chun-yang.chen@tum.de}

\renewcommand{\shortauthors}{Anonymous authors}

\begin{abstract}
Large Language Models (LLMs) have revolutionized human-AI interaction by enabling intuitive task execution through natural language prompts. Despite their potential, designing effective prompts remains a significant challenge, as small variations in structure or wording can result in substantial differences in output. To address these challenges, LLM-powered applications (\llmapps) rely on prompt templates to simplify interactions, enhance usability, and support specialized tasks such as document analysis, creative content generation, and code synthesis.
However, current practices heavily depend on individual expertise and iterative trial-and-error processes, underscoring the need for systematic methods to optimize prompt template design in \llmapps. This paper presents a comprehensive analysis of prompt templates in practical \llmapps. We construct a dataset of real-world templates from open-source \llmapps, including those from leading companies like Uber and Microsoft. Through a combination of LLM-driven analysis and human review, we categorize template components and placeholders, analyze their distributions, and identify frequent co-occurrence patterns. Additionally, we evaluate the impact of identified patterns on LLMs' instruction-following performance through sample testing. Our findings provide practical insights on prompt template design for developers, supporting the broader adoption and optimization of \llmapps\ in industrial settings.
\end{abstract}




\received{xxx}
\received[revised]{xxx}
\received[accepted]{xxx}

\maketitle

\section{Introduction}
Large Language Models (LLMs), such as GPT-4~\cite{achiam2023gpt} and LLaMA~\cite{touvron2023llama}, have exhibited exceptional capabilities in comprehending and generating text, and LLMs have rapidly evolved to become a cornerstone of many AI-driven applications. This versatility has introduced a new paradigm of interaction between humans and AI, where users express their instructions in natural language, and the LLM generates outputs that align with these specifications. At first glance, a prompt may appear to be a simple question or request, but it actually serves as a bridge between human intent and machine-generated responses, guiding the model by providing context, shaping outputs, and influencing behavior.
While these models have significantly reduced the barriers to AI adoption for general users, crafting clear and effective prompts remains a non-trivial challenge~\cite{Desmond2024ExploringPE, ZamfirescuPereira2023WhyJC}. This difficulty arises from the limited understanding of how LLMs process input information and the fact that even minor variations in prompts can lead to substantial changes in model performance~\cite{salinas2024butterfly, sclar2023quantifying}.

To address these challenges, numerous \llmapps, which refer to the type of software that uses LLMs as one of its building blocks, have been developed~\cite{yan2024exploring, hou2024security} in different downstream domains.  
The growth of \llmapps\ has been remarkable, with over one million applications released for public use and leading \llmapps\ now boast more than three million monthly active users~\cite{zhao2024llm}. 
For example, ``Write For Me'' is one of the most popular \llmapps\ in the GPT store, with over 12 million conversions and a high rating of 4.3 based on over 10,000 reviews~\cite{write_for_me_gpt}.
Unlike traditional software development that involves full-stack implementation, \llmapps\ adopt a low-code paradigm where LLMs handle most user's queries, and developers focus on facilitating user-LLM interaction by designing lightweight prompt templates which is a predefined structure that combines static text with dynamic placeholders, allowing developers to create adaptable prompts for \llmapps ~\cite{gemini_prompt_templates}. These predefined prompt templates simplify and standardize user interactions with LLMs~\cite{gorissen2024survey, schulhoff2024prompt}, ensuring consistency and efficiency across various tasks.

\begin{figure*}
    \centering
\includegraphics[width=\linewidth]{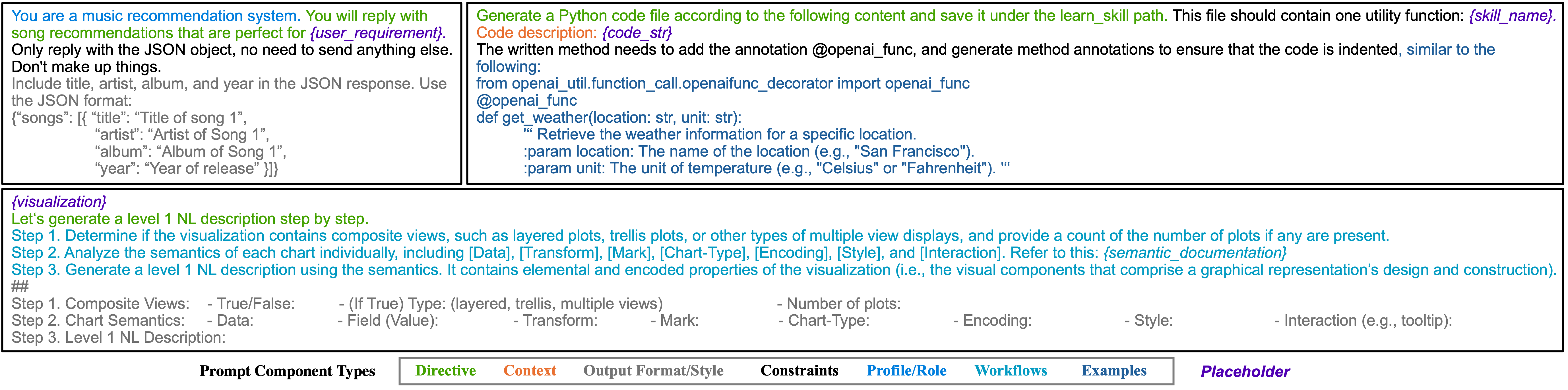}
    \caption{Examples of prompt template~\cite{beets_plexsync,Andrew,chart-llm}} 
    \label{fig:prompt_template_example}
    \vspace{-3mm}
\end{figure*}

Figure~\ref{fig:prompt_template_example} presents examples of prompt templates designed for various tasks, showcasing their components and placeholders. For example, the first template establishes the model’s role, provides task-specific directives (\eg, song suggestion), imposes constraints to reduce model hallucination, and defines the expected output format to be a JSON string with specific attributes. The \textit{\{user\_requirement\}} placeholder allows for diverse user inputs, such as ``light music for bedtime'' or ``a song with a summer feel describing ... life in ...'' The other two templates demonstrate the use of examples and detailed workflow descriptions, employing in-context learning and chain-of-thought techniques to guide the LLMs effectively.
Serving as critical artifacts in \llmapps, prompt templates reduce the complexity of crafting effective prompts, bridging the gap between user intent and machine responses. 
Similar to GUI (Graphical User Interface) which bridges the gap between backend programs and end-users, prompt templates are like a ``textual GUI'' for AI systems, where inputs follow a structured format requiring only basic inputs, such as a few clicks or filling in the blank, instead of long complete instruction input.
Just as GUIs have widgets that can be reused or customized, prompt templates have components and placeholders that allow flexibility and dynamic usage which hide the intricate operations of the backend from the user.
From setting the context and defining instructions to dynamically adjusting the content based on user needs, prompt templates offer a versatile approach to language model interactions.
By providing a consistent structure, prompt templates help users articulate their requirements more effectively, ensuring that the LLM understands and responds appropriately even in specialized domains, such as document analysis~\cite{wang2024knowledge}, creative content generation~\cite{chen2024benchmarking}, and code synthesis~\cite{guo2024stop}. 

As a new paradigm to develop \llmapps\, almost all commercial LLM providers support prompt templates, such as OpenAI's GPT series~\cite{openai}, Google's Gemini~\cite{gemini_prompt_templates}, Anthropic's Claude~\cite{anthropic_prompt_template}, Salesforce~\cite{salesforce_prompt_templates}, and Langchain~\cite{langchain_prompt_templates}. Unlike traditional software development, which is deterministic and relies on well-defined code~\cite{li2023cctest, holter2024abstract, yuan2024evaluating}, \llmapps\ introduce variability in outputs, challenging software engineers to design prompts that ensure reliability, maintainability, and performance.
However, it is still unclear how to develop high-quality prompt templates which are not even clearly mentioned in the documentation from these leading commercial LLM providers.
What information should developers put into prompt templates? What patterns do they follow? What dynamic content will prompt templates take from end users? How do different construction patterns in prompt templates influence LLMs' instruction-following abilities?
Considering the growing significance of prompt templates in the \llmapp\ ecosystem, a systematic understanding of their structure, composition, and effectiveness is urgently needed.

In this paper, we analyze prompt templates used in LLM-powered applications, focusing on their components, placeholders, and composition patterns. Our study utilizes a publicly available dataset of prompts collected from open-source \llmapps~\cite{pister2024promptset}, encompassing applications from IT giants to innovative startups that reflect real-world use cases and industry trends, such as tools from Uber (adopted by over 200 developers)~\cite{ramanathan2020piranha, ketkar2024lightweight} and Microsoft (over 5k GitHub stars)~\cite{qiao2023taskweaver}.
We pre-process the dataset to filter representative prompt templates and systematically identify their components and placeholders. By analyzing the content and order of these elements, we uncover structural and content patterns commonly employed in prompt template design. To evaluate these patterns, we conduct sample testing using randomly selected templates, assessing their impact on LLM instruction-following abilities and identifying optimal patterns for enhanced performance.

Our contributions are summarized as follows:
\begin{itemize}[topsep=0pt,leftmargin=0.7cm]
    \item To the best of our knowledge, this is the first study analyzing real-world prompt templates in \llmapps, with datasets publicly available at: \href{https://github.com/RedSmallPanda/FSE2025}{https://github.com/RedSmallPanda/FSE2025}.
    \item We categorize key components, placeholders, and patterns from collected large-scale prompt templates, offering insights into effective prompt template design practices. 
    \item Through sample testing, we demonstrate that well-structured prompt templates and specific composition patterns can significantly improve LLMs' instruction-following abilities.
\end{itemize}

\section{Methodology}

\subsection{Overview}

\begin{figure}[t]
 \centering
  \includegraphics[width=\linewidth]{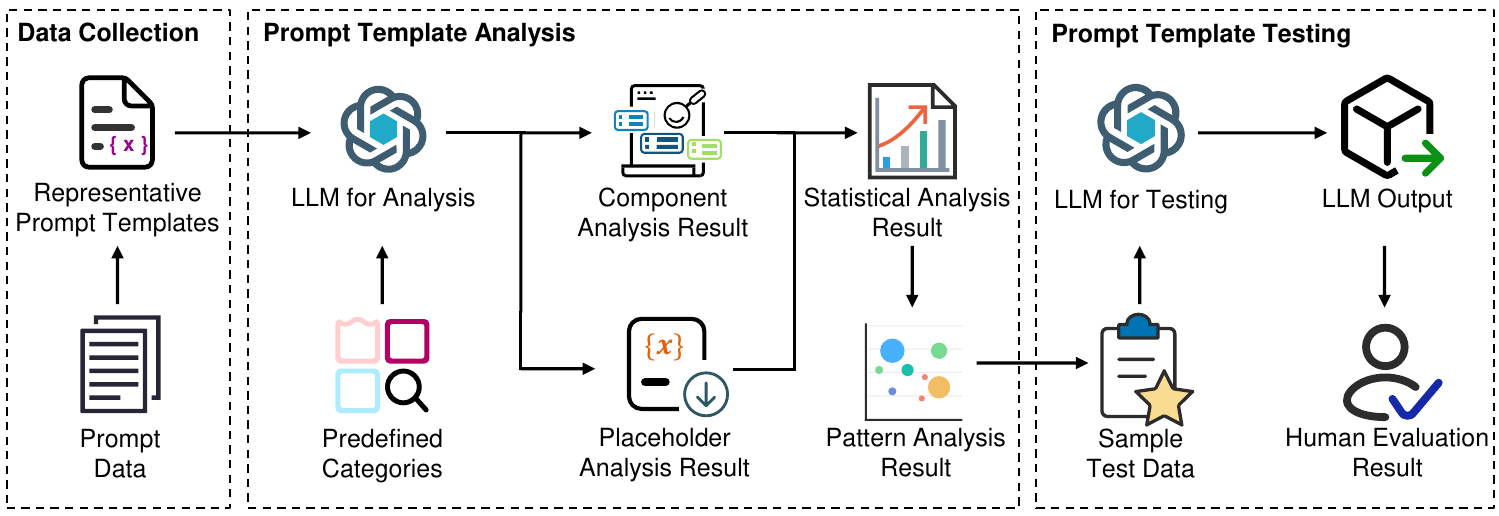}
  \caption{An illustration of the main pipeline}
  \vspace{-5mm}
  \label{fig:overview}
\end{figure}

Figure~\ref{fig:overview} shows the main framework of our method including three main steps: the construction of large-scale prompt templates from real-world GitHub repositories, an empirical study of prompt templates focusing on components, placeholders, patterns, and an exploratory study to check the effects of these prompt template patterns to LLM output.

\subsection{Data Collection}


We construct our dataset based on PromptSet~\cite{pister2024promptset}, a collection of prompts extracted from \llmapps in open-source GitHub projects as of January 10, 2024. These projects vary widely in complexity, usage, and popularity, ranging from personal demos to widely adopted applications, resulting in significant variability in prompt quality. To ensure a high-quality dataset suitable for industrial applications, we design a processing pipeline that assigns quality metrics to each prompt and automatically filters out lower-quality examples.

Our data collection pipeline begins by selecting non-empty English prompts from the PromptSet dataset, resulting in 14,834 records. For each corresponding GitHub repository, we retrieve metadata such as \textit{star count} and \textit{latest update time} to evaluate repository popularity and activity~\cite{weber2014makes, cosentino2016findings}, utilizing the GitHub API, accessed on June 20, 2024.
We filter repositories with at least five stars and recent updates within the past year, narrowing the dataset to 2,888 records across 1,525 repositories. Next, we separate multi-prompt records into individual entries, resulting in 5,816 prompts, and then remove duplicates to obtain 4,540 unique prompts. To further enhance data quality, we exclude prompts shorter than five tokens\cite{spacy2}, with 4,210 high-quality prompts left. Finally, we extract 2,163 distinct prompt templates using the \llamaversion\ model, guided by a clear definition from Schulhoff~\etal~\cite{schulhoff2024prompt}, supplemented by illustrative examples. 
Most repositories in our dataset are \llmapps\ serving for information seeking, editing, coding, and creative writing.
As seen in Table~\ref{tab:repo}, some of them are from leading companies and organizations, such as Uber's tool for refactoring code related to feature flag APIs (adopted by over 200 developers~\cite{ramanathan2020piranha, ketkar2024lightweight}), Microsoft’s code-first agent framework for executing data analytics tasks with over 5k GitHub stars~\cite{qiao2023taskweaver} and some very popular ones, such as an RAG chatbot powered by Weaviate with over 6k GitHub stars~\cite{verba}, and LAION-AI's Open-Assistant program, with more than 37k stars~\cite{open_assistant}.

\begin{table}[t]
\centering
\caption{Representative \llmapp\ repositories~\cite{ketkar2024lightweight, qiao2023taskweaver, verba, open_assistant}}
\resizebox{\linewidth}{!}{
    \label{tab:repo}
    \begin{tabular}{l l l l }
        \toprule
        \textbf{Company Name} & \textbf{Repo Title} &  \textbf{Description}  & \textbf{\#Stars} \\
        \midrule
       Uber & uber/piranha & Tool for refactoring code related to feature flag APIs  & 2.3k \\
      Microsoft & microsoft/TaskWeaver & Code-first agent framework for data analytics  & 5.4k  \\
      Weaviate & weaviate/Verba  & RAG chatbot & 6.5k\\
      LAION & LAION-AI/Open-Assistant & Chat-based assistant & 37.1k\\
        \bottomrule
    \end{tabular}
}
\begin{tablenotes}
\footnotesize
\item \#Stars assessed on 30.12.2024.
\end{tablenotes}
\end{table}

\subsection{Prompt Template Analysis}

In this section, we address the following research questions:
\begin{itemize}[leftmargin=0.4cm]
 \item \textbf{RQ1: What are the common constituent components in prompt templates, and what specific words, phrases, and patterns are frequently used within these components?}
    
    We investigate the component composition of representative prompts extracted from \llmapps\ and analyze co-occurrence patterns among components. Building on these findings, we conduct word-level and phrase-level analysis of each component type to uncover variations in how similar content is expressed (\eg, different ways users describe the output format as ``JSON''). From these variations, we identify common patterns that impact response format and content.

    \item \textbf{RQ2: How do placeholders vary in terms of types and positions within prompt templates?}
    
    We analyze the types of placeholders (variables that will be replaced by input text to create a complete prompt) within prompt templates~\cite{schulhoff2024prompt}, categorizing them based on their variable names and contextual usage within the templates. Then we examine the positional distribution of placeholders across prompt templates.
\end{itemize}

\subsubsection{Prompt Component Analysis}\ 

\noindent In this step, we define a comprehensive set of common prompt components by synthesizing insights from prominent prompt engineering frameworks and AI service platforms. Specifically, we extract component definitions from guidelines provided by Google Cloud's documentation~\cite{GoogleCloud}, the Elavis Saravia framework~\cite{saravia2022prompt}, the CRISPE framework~\cite{ChatGPT3FreePromptList}, and the LangGPT framework~\cite{Wang2024LangGPTRS}. To construct a unified list, we merge similar components across these sources. For instance, components like Profile, Capacity, Role, and Persona, all of which define the model's behavior or identity, are consolidated into a single component labeled ``Profile/Role.'' 
The merged component list is presented in Table~\ref{tab:merge}, and component definitions and frequency distributions are detailed in Table~\ref{tab:component_distribution}.

\begin{table}[htpb]
  \caption{Prompt components across different frameworks and documentations.}
  \centering
 \resizebox{\linewidth}{!}{
  \begin{tabular}{l l l l l }
    \toprule
    \textbf{LangGPT \cite{Wang2024LangGPTRS}} & \textbf{Elavis Saravia \cite{saravia2022prompt}} & \textbf{CRISPE \cite{ChatGPT3FreePromptList}} & \textbf{Google Cloud\cite{GoogleCloud}}   & \textbf{Merged}\\ 
    \midrule
    Profile & - & Capacity and Role & Persona  & \textbf{Profile/Role} \\
    \midrule
    \multirow{3}{*}{Goal} & \multirow{3}{*}{Instruction} & \multirow{3}{*}{Statement} & Objective & \multirow{3}{*}{\textbf{Directive}}\\
    & & & Instructions &\\
    & & & System Instructions  &\\
    \midrule
    Workflow & - & - & Reasoning Steps  & \textbf{Workflow} \\
    \midrule
   Initialization& \multirow{2}{*}{Context} & \multirow{2}{*}{Insights} & \multirow{2}{*}{Context}  & \multirow{2}{*}{\textbf{Context}}\\ 
   Background  & &  &     \\
    \midrule
    Example & Input Data & - & Few-shot Examples   & \textbf{Examples}\\ 
    \midrule
    Output-format & \multirow{2}{*}{Output Indicator} & \multirow{2}{*}{Personality} & Response Format  & \multirow{2}{*}{\textbf{Output Format/Style}}\\ 
    Style &  &  & Tone  & \\
    \midrule
    \multirow{2}{*}{Constraints}& \multirow{2}{*}{-} & \multirow{2}{*}{-} & Constraints  &  \multirow{2}{*}{\textbf{Constraints}} \\ 
     & & & Safeguards & \\
    \midrule
    Skill & \multirow{2}{*}{-} & \multirow{2}{*}{Experiment} & \multirow{2}{*}{Recap}  &  \multirow{2}{*}{\textbf{Others}} \\
    Suggestion &  &  & \\ 
    \bottomrule
  \end{tabular}
  }
  \label{tab:merge}
\end{table}

We leverage the \llamaversion\ model to identify components from the merged list present in the prompts, employing a predefined prompt template that specifies all available components and outputs results in a structured JSON format. 

To assess the accuracy of component identification, we perform both component-level and prompt-level human evaluations on a randomly selected 5\% sample of prompts. At the component level, precision is calculated as the proportion of correctly identified components compared to human-labeled ones. At the prompt level, a prompt is classified as an exact match only if all identified components are correct; prompts with at least one correct component are classified as partial matches~\cite{katsogiannis2023survey}.

Human evaluations follow established practices~\cite{li2023skcoder}, with two evaluators both with over one year of programming experience in \llmapp\ independently reviewing LLM-generated classifications. A component is considered correctly identified if both evaluators agree. Placeholder identification accuracy is validated in the same manner. For prompt template testing, evaluators score LLM outputs based on predefined metrics, with the final score being the average of their assessments.

The evaluation results indicate high component-level precision, averaging 86\% across all predefined types. At the prompt level, full match precision is 66\%, while partial match precision reaches 99\%, demonstrating the method's reliability in component detection.




\subsubsection{Placeholder Analysis}\ 

\noindent We analyze placeholders within prompt templates using a two-step iterative process. First, we manually classify placeholders from a randomly selected set of 100 templates into predefined or newly identified categories based on their names, following practices similar to variable analysis~\cite{avidan2017effects}. Then, we use \gptversion\ to extend this classification to the full dataset, leveraging the initial categories and definitions.

To ensure accuracy, we conduct human evaluations to verify the LLM's classifications, merging underrepresented categories and refining definitions as needed. After incorporating these adjustments, we perform a second round of LLM classification, followed by another human evaluation. This iterative process achieves an overall classification accuracy of 81\% on a randomly selected sample of 80 records. Ultimately, we identify four primary placeholder categories, as detailed in Table~\ref{tab:placeholder_type_distribution}.


\subsubsection{Statistical Analysis and Pattern Analysis}\ 

\noindent Using the component analysis and placeholder analysis results, we conduct a statistical analysis to examine the distribution of each category, as well as the frequency and relative positioning of elements at the word and phrase levels. From these statistical findings, we extract notable patterns that reveal structural and content trends within the prompt templates.

\subsection{Effects of Prompt Template Patterns}
In this section, we address the following research question:
\begin{itemize}[leftmargin=0.4cm]
    \item \textbf{RQ3: How do different construction patterns in prompt templates influence LLMs' instruction-following abilities?}
    
    We sample prompt templates with specific patterns identified in earlier research questions, populate them with sample data, and generate outputs to assess the effectiveness of these patterns on LLMs' instruction-following abilities through human evaluation.
\end{itemize}

In this phase, we assess the impact of various patterns identified in the analysis of RQ1 and RQ2 on LLM output, focusing on two key dimensions: Content-Following, which ensures semantic accuracy with task goals, and Format-Following, which enforces adherence to structural or syntactical requirements critical for \llmapps\ but underexplored in prior research~\cite{xia2024fofo}. Specific metrics are defined for each test to evaluate one or both dimensions.

For each pattern, we randomly sample prompt templates that incorporate it and populate them using either the \gptversion\ model or real-world data sources, such as a medical QA dataset~\cite{ben2019question}, Java documentation~\cite{jdk22}, and GitHub projects~\cite{transformer}. To compare patterns addressing similar issues, we manually reformulate the templates (\eg, modifying the output format or adding constraints) to fit alternative patterns. Outputs are then generated using both \llamaversion\ and \gptversion\ models, followed by human evaluations to assess output quality against predefined metrics.

\section{Analysis Results}

\subsection{RQ1: Analyzing Components in Prompt Templates}
\begin{figure}[]
    \centering
    \scriptsize 
    \renewcommand{\arraystretch}{0.55}
        \begin{tabular}{|@{\hskip 5pt}p{0.78\linewidth}@{\hskip 5pt}|p{0.14\linewidth}@{}}
            \cline{1-1}
            \\
            \textcolor{TUMBlue}{You are an AI assistant acting as a content advisor for a tech blog.}  
            \textcolor{TUMAccentGreen}{Suggest two relevant blog topics based on recent trends in \textbf{\{subject\_area\}}.} 
            & \textcolor{TUMBlue}{\textbf{Profile/Role}} \textcolor{TUMAccentGreen}{\textbf{Directive}}  \\
            \\
            \textcolor{TUMAccentOrange}{You are given the following information:} & \textcolor{TUMAccentOrange}{\textbf{Context}} \\
            \textcolor{TUMAccentOrange}{- The blog focuses on topics related to \textbf{\{subject\_area\}.}} &  \\
            \textcolor{TUMAccentOrange}{- Recent posts on \textbf{\{high\_engagement\_topics\}} have received high engagement.} &  \\
            \\
            \begin{minipage}[t]{\linewidth}
            \textcolor{TUMAccentBlue}{To complete this task, follow these steps:}\\
                \textcolor{TUMAccentBlue}{1. Review the provided context and analyze current trends in \textbf{\{subject\_area\}}.} \\
                \textcolor{TUMAccentBlue}{2. Suggest two blog topics that align with the blog’s focus and audience.} \\
                \textcolor{TUMAccentBlue}{3. Ensure the suggested topics are relevant to the recent engagement trends, particularly \textbf{\{high\_engagement\_topics\}}, and are accessible to a general audience.}
            \end{minipage} & \textcolor{TUMAccentBlue}{\textbf{Workflows}} \\
            \\
            \textcolor{TUMDarkGray}{Avoid overly technical or niche topics that may overwhelm readers. Keep topics broad and engaging for non-experts.} & \textcolor{TUMDarkGray}{\textbf{Constraints}} \\
            \\
            \begin{minipage}[t]{\linewidth}
                \textcolor{TUMGray}{Provide your response in the following format:} \\
                \textcolor{TUMGray}{Topic 1: [Title] - [One-sentence explanation]} \\
                \textcolor{TUMGray}{Topic 2: [Title] - [One-sentence explanation]}
            \end{minipage} & \textcolor{TUMGray}{\textbf{Output Format/Style}} \\
            \\
            \begin{minipage}[t]{\linewidth}
            \textcolor{TUMSecondaryBlue}{There are two example topics based on trends in AI:} \\
                \textcolor{TUMSecondaryBlue}{Topic 1: AI in Public Services - Discuss how AI is being used to enhance efficiency in public services like healthcare and education.} \\
                \textcolor{TUMSecondaryBlue}{Topic 2: Ethical AI in Automation - Explore the ethical implications of AI-driven automation in industries like manufacturing and logistics.}
            \end{minipage} &  \textcolor{TUMSecondaryBlue}{\textbf{Examples}} \\
            \\
            \cline{1-1}
        \end{tabular}
        \label{fig:pattern_a}
    \caption{A Template Example following General Component Order.}
    \vspace{-3mm}
    \label{fig:pattern_all_elements}
\end{figure}

\subsubsection{Distribution of Components}\ 

\noindent Table~\ref{tab:component_distribution} shows the detection results for the seven categories of components, indicating the percentage of prompt templates containing each component.
Among these components, the four most common are Directive, Context, Output Format/Style, and Constraints. The Directive represents the task intent of the prompt, guiding the language model on how to perform a task. Most prompts require a clear and complete directive to instruct the model effectively. The Context typically includes the input content and relevant contextual descriptions, helping the model understand the task in detail. Given that these prompt templates are designed for \llmapps, developers often specify an Output Format/Style (\eg, Topic: Title - Explanation, as illustrated in Figure~\ref{fig:pattern_all_elements}) and set Constraints (\eg, length, number of results, output topic scope). This ensures the generated content is more predictable and easier for downstream applications to process, and maintains consistency across outputs.
\begin{table}[t]
  \caption{Frequencies distribution of different prompt components.}
  \centering
 \resizebox{\linewidth}{!}{
  \begin{tabular}{l p{0.7\linewidth}  l  }
    \toprule
    \textbf{Components} & \textbf{Definition} & \textbf{Frequency}\\
    \midrule
    Profile/Role & Who or what the model is acting as. & 28.4\%\\
   \rowcolor{gray!10} Directive & The core intent of the prompt, often in the form of an instruction or a question. & 86.7\%\\
    Workflow & Steps and processes the model should follow to complete the task. & 27.5\% \\
   \rowcolor{gray!10} Context & Background information and context that the model needs to refer to. & 56.2\%\\
    Examples & Examples of what the response should look like. & 19.9\%\\
   \rowcolor{gray!10} Output Format/Style &  The type, format, or style of the output. & 39.7\%\\
    Constraints & Restrictions on what the model must adhere to when generating a response. & 35.7\%\\
    \bottomrule
  \end{tabular}
  }
  \label{tab:component_distribution}
\end{table}

\begin{figure}[t]
    \centering     \includegraphics[width=0.8\linewidth]{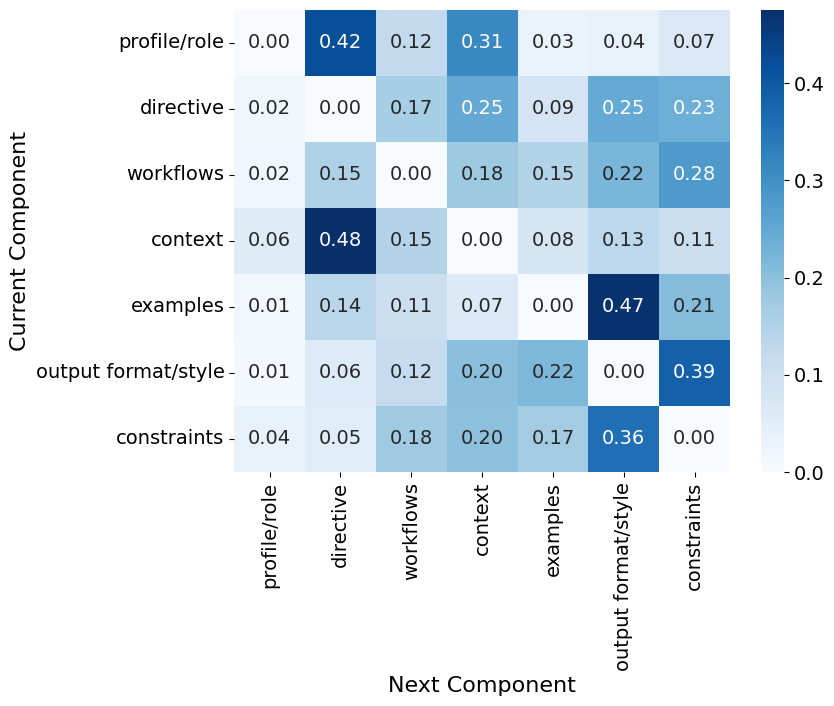} 
    \caption{Component order probability matrix}
    \vspace{-3mm}
    \label{fig:element_transition}
\end{figure}

\subsubsection{Component Order}\label{section:Component Order}\

\noindent We also investigate the relative positions of various components. We observe that the Profile/Role and Directive components commonly appear in the first position rather than other positions, with a probability of 0.87 and 0.65, respectively. As shown in Figure~\ref{fig:element_transition}, both the X and Y axes represent the different component types, with each coordinate point indicating the probability that component X follows component Y. The darker the color, the higher the probability that this positional pattern occurs. For example, Context is most likely to be followed by Directive with a probability of 0.48. 

\begin{figure}[t]
    \centering
      \includegraphics[width=\linewidth]{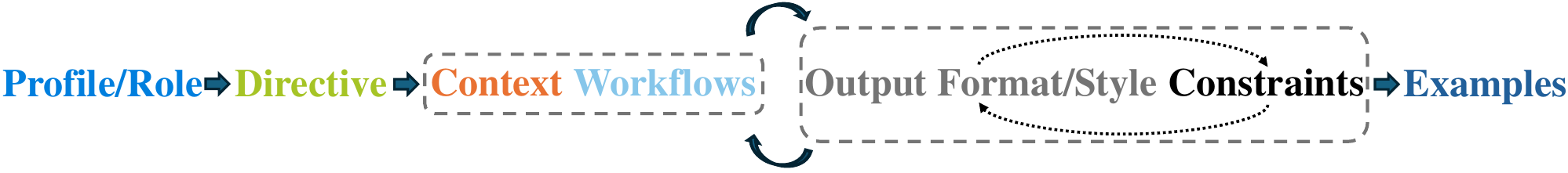}
    \caption{A Common Component order}
    \vspace{-3mm}
    \label{fig:component_order}
\end{figure}

Based on the analysis, we identify a common sequential order of components in prompt templates, as depicted in Figure~\ref{fig:component_order}. This sequence assumes each component appears exactly once. Notably, ``Context and Workflows'' and ``Output Format/Style and Constraints'' consistently form two pairs, where the order within the ``Output Format/Style and Constraints'' pair is flexible. Additionally, the relative positioning of these two pairs can also be interchanged without impacting overall structure. Figure~\ref{fig:pattern_all_elements} illustrates a specific prompt example that adheres to this identified order.

\begin{tcolorbox}[colback=black!5!white,colframe=black!75!black,before upper=\vspace{-.25cm},after upper=\vspace{-.25cm},left=0.1pt, right=0.1pt]
  \textbf{Finding 1: } There are mainly seven types of common components within prompt templates according to our observation. Developers commonly put the profile/role and their directive at the beginning of the prompt, establishing the model’s identity and task intent, while examples are typically at the end.  
\end{tcolorbox}

\subsubsection{Component Content}\label{section:component_content}\ 

\noindent We perform an in-depth analysis of specific prompt components that have a significant impact on the structure consistency and instruction relevance of LLM responses, focusing on commonly used words, phrases, and formats associated with these components. Among the seven types of prompt components, we focus on three: Directive, Output Format/Style and Constraints. 
In direct interactions with LLMs (such as live chat), users primarily care about the relevance and correctness of the response. However, in \llmapps, developers are not only concerned with the response content but also with its format, as post-processing often relies on structured outputs. Ensuring a limited or predictable output format can significantly reduce errors during post-processing.
Based on our observations of how \llmapps\ handle responses, we focus on Directive, which encapsulates the core user intent, along with Output Format/Style and Constraints, as they play a critical role in shaping both response format and content—key factors for downstream processing and application performance.

\noindent \textbf{Directive. } Directive in prompts could be either in question or instruction style~\cite{schulhoff2024prompt, Karmaker2023TELeRAG}. Using regex patterns to capture directives starting with question words such as ``how'' and ``what'' or ending with a question mark ``?'', we classify the directive into two types, namely instruction and question. Notably, our analysis shows that over 90\% of directives are written in the instruction style. This is likely due to the fact that instructions such as ``Summarize the report'' are more direct and clearer for the model to understand than question-style directives like ``Could you summarize this?''.

\begin{tcolorbox}[colback=black!5!white,colframe=black!75!black,before upper=\vspace{-.25cm},after upper=\vspace{-.25cm},left=0.1pt, right=0.1pt]
  \textbf{Finding 2: } The directive component i.e., user intent in prompt templates predominantly adopts an instructional style, which is more commonly used than the question style.
\end{tcolorbox}

\begin{figure}[t]
    \centering
    \subfloat[][Information Seeking.]{
    \includegraphics[width=0.3\linewidth]{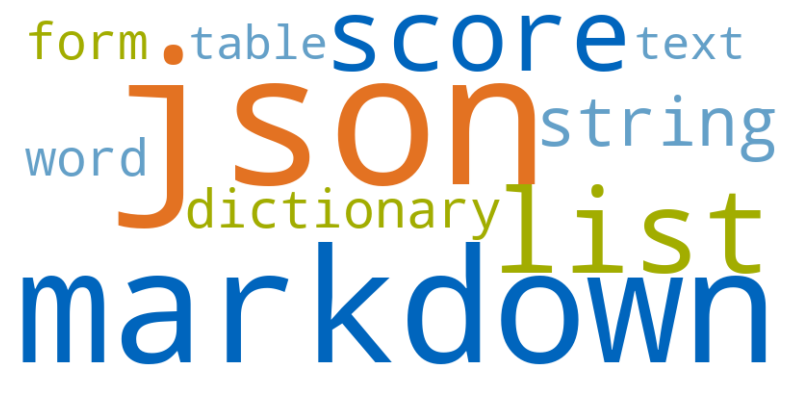}\label{fig:format_info}}
    \hspace{0.01\linewidth}
    \subfloat[][Coding \& Debugging.]{ \includegraphics[width=0.3\linewidth]{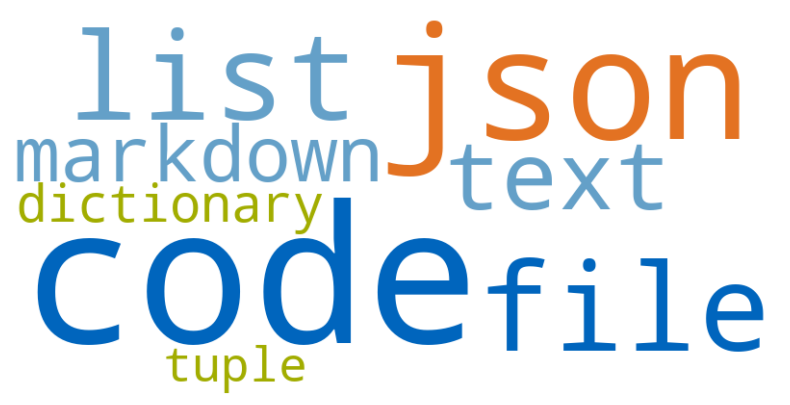}\label{fig:format_coding}}
    \subfloat[][Editing.]{ \includegraphics[width=0.3\linewidth]{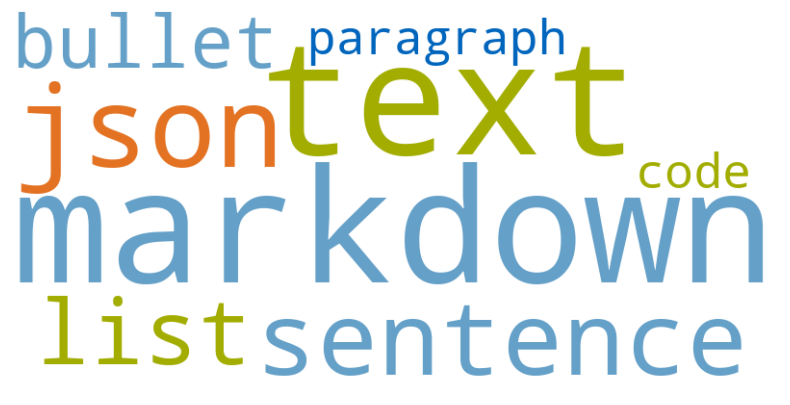}}\label{fig:format_editing}
    \caption[]{Word cloud for output format of selected themes}
    \vspace{-5mm}
    \label{fig:output_format_word_cloud}
\end{figure}

\noindent \textbf{Output Format/Style. }We analyze the output formats specified in the prompt templates across different themes. To extract these formats, we consider the most frequently occurring terms in the output descriptions of the prompt templates and map them into word clouds for each theme, as shown in Figure~\ref{fig:output_format_word_cloud}. Words in the cloud are sized proportionally to their frequency: the larger the word, the more often that output format is used within the theme.

From our analysis, certain formats like ``score'' and ``code'' are frequently used in particular themes. For example, ``code'' is a predominant output format in the ``Coding \& Debugging'' tasks, often appearing as the result of tasks such as code generation, bug fixing, and code summarization. Similarly, ``score'' appears frequently in ``Information Seeking'' tasks, where users provide criteria for LLMs to evaluate inputs and generate numerical scores.

Across all themes, the most common output format besides standard text is JSON. JSON's structured nature makes it particularly popular, as it is easy for applications to post-process and provides a user-friendly way to organize complex information. 

\begin{figure}[t]
    \centering
      \includegraphics[width=\linewidth]{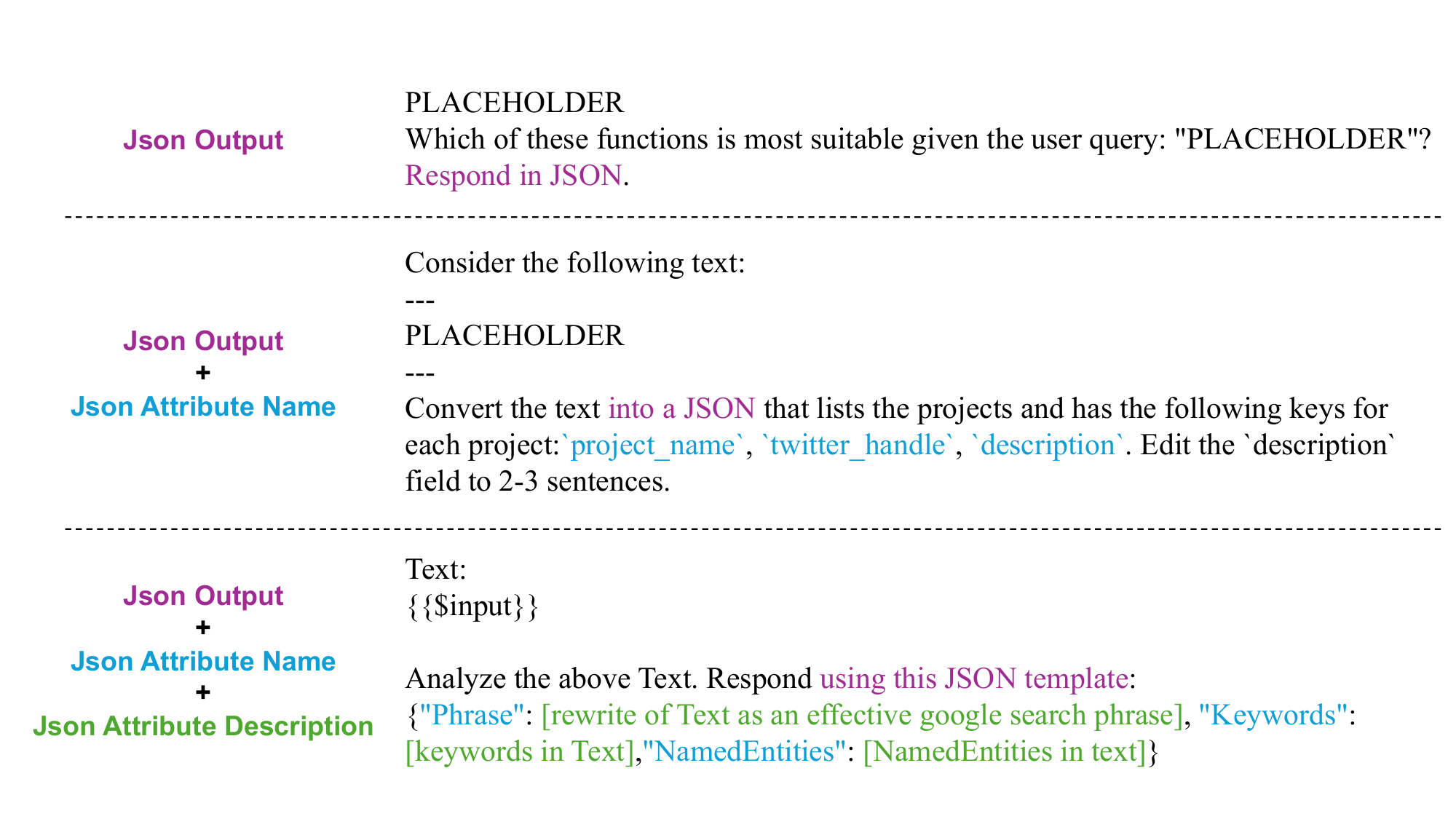}
    \caption{Three different JSON output formats~\cite{definitive,hypercerts,agentm-py}}
    \vspace{-3mm}
    \label{fig:json_format}
\end{figure}

To analyze how developers define JSON output formats, we extract all prompt templates that specify JSON as the required output format. From the data, we identify three key components frequently used: \textit{Json Output}, which specifies that the output must be in JSON format; \textit{Json Attribute Name}, which defines the names of the attributes to be included; and \textit{Json Attribute Description}, which provides detailed explanations for each attribute. These components combine into three distinct patterns. In \textit{Pattern 1}, developers only indicate that the output should be in JSON format, often accompanied by general guidelines in natural language about the expected content. \textit{Pattern 2} builds on this by explicitly listing the attribute names to be included, ensuring structural consistency in the output. Finally, \textit{Pattern 3} adds detailed descriptions for each attribute to Pattern 2, enhancing clarity and ensuring attributes are well-defined, particularly for complex outputs. Figure~\ref{fig:json_format} illustrates these patterns and their composition. 



The distribution of these patterns across all observed templates reveals that Pattern 1 (Json Output) accounts for 36.21\%, Pattern 2 (Json Output + Json Attribute Name) for 19.83\%,
and Pattern 3 (Json Output + Json Attribute Name + Json Attribute Description) for 43.97\%.

\begin{tcolorbox}[colback=black!5!white,colframe=black!75!black,before upper=\vspace{-.25cm},after upper=\vspace{-.25cm},left=0.1pt, right=0.1pt]
  \textbf{Finding 3: } JSON is the most commonly used output format in prompt templates. 
  Notably, over one-third of these JSON formats rely on informal descriptions without explicitly defined attribute names.
\end{tcolorbox}

\noindent \textbf{Constraints.} We use the \llamaversion\ model to identify various constraint types in prompt templates, based on the classifications established by Ross Dawson~\cite{rossdawson_prompt_elements}. 
The most common types are ``Exclusion'' (46.0\%), ``Inclusion'' (35.6\%), and ``Word count'' (10.5\%). These categories reflect the primary considerations developers emphasize when designing prompt templates. The ``Inclusion'' constraints guide the model to focus on specific details, improving the relevance and precision of its responses. ``Word count'' constraints encourage concise responses, enhancing usability and reducing API costs by minimizing token usage.
\begin{table*}[htpb]
\caption{Examples of exclusion cluster}
\centering
\tiny
\resizebox{\textwidth}{!}{
\begin{tabular}{p{2.7cm} p{5.0cm} p{4.7cm}}
\toprule
\textbf{Exclusion Cluster} & \textbf{Exclusion Examples} & \textbf{Complementary Inclusion Examples}\\
\midrule
\multicolumn{3}{c}{\textbf{Accuracy Output}} \\ 
\midrule

Accuracy and relevance &  
- avoid adding any extraneous information. & - include only crucial information relevant to...\\  \midrule

Clarity about unknowns & 
- if you don't know the answer, just say that you don't know, don't try to make up an answer.
 & - if you don’t know the answer, you may make inferences, but
make it clear in your answer.\\
\midrule
\multicolumn{3}{c}{\textbf{Concise Output}} \\ 
\midrule

Output control (what text/code should be excluded) & 
- do not provide any other output text or explanation. \newline 
- if you are calling a function, only include the function call -- no other text. & - time information should be included \newline - include at most 10 of the most related links.\\ 
\midrule

Redundancy and context adherence & 
- don't give information outside the document or repeat your findings. \newline 
- do not generate redundant information. \\  \midrule

\multicolumn{3}{c}{\textbf{Technical Restriction}} \\  \midrule

 Technical restriction & 
- never write any query other than a select, no matter what other information is provided in this request. & - obey the ros package name conventions when choosing the name.\\ 
\bottomrule

\end{tabular}
}
\label{tab:not_include}
\end{table*}

Given that ``Exclusion'' constraints account for nearly half of all identified constraints, we conducted a detailed analysis of this category. Using the all-mpnet-base-v2 embedding model and k-means clustering, we identified five subcategories: ``Accuracy and Relevance'', ``Clarity about Unknowns'', ``Output Control'', ``Redundancy and Context Adherence'', and ``Technical Restriction'' (Table~\ref{tab:not_include}). These subcategories reflect nuanced developer intent in managing model outputs.

Key subcategories, such as ``Accuracy and relevance'' and ``Clarity about unknowns'' are designed to mitigate hallucinations, a prevalent issue with LLMs. For example, prompts like ``Don’t try to make up an answer'' reduce the risk of speculative or incorrect outputs, enhancing reliability. Anthropic's documentation~\cite{anthropic_hallucination_2024} and Chen \etal~\cite{Chen2024HonestAF} propose similar constraints to mitigate the hallucination.
Additionally, constraints targeting redundancy, such as ``Do not generate redundant information'' streamline responses, reducing post-processing needs in tasks like code generation. The ``Technical restriction'' subcategory highlights constraints for specific contexts, such as database queries or API calls, ensuring that models adhere to precise technical requirements like predefined structures or restricted API usage. These constraints are critical for maintaining correctness and efficiency in structured workflows.

\begin{tcolorbox}[colback=black!5!white,colframe=black!75!black,before upper=\vspace{-.25cm},after upper=\vspace{-.25cm},left=0.1pt, right=0.1pt]
  \textbf{Finding 4: } Developers frequently use exclusions as the constraints to refine outputs such as excluding irrelevant content, reducing hallucinations, and narrowing search space for generation.
\end{tcolorbox}

\subsection{RQ2: Analyzing Placeholders in Prompt Templates}

\subsubsection{Classification of Placeholder}\ 

\noindent Table~\ref{tab:placeholder_type_distribution} displays the percentage of prompt templates containing each placeholder category. The most prevalent categories are Knowledge Input and Metadata/Short Phrases. Knowledge Input placeholders contain the main content with which the LLM directly interacts, including items like ``report'' or ``code snippet.'' Metadata/Short Phrases serve as brief inputs providing essential settings or specific details, such as ``language” or ``username.'' Developers tend to include multiple Metadata/Short Phrases within one template. Additionally, the User Question placeholder captures the direct query from the end user, while Contextual Information placeholders provide supplementary background content, such as “chat history” or “background info,” offering context that supports the task without being central to it.

\begin{table}[t]
\centering
\caption{Distribution of placeholder type in prompt templates.}
\resizebox{\linewidth}{!}{
    \label{tab:placeholder_type_distribution}
    \begin{tabular}{l p{0.5\linewidth} p{0.3\linewidth} l}
        \toprule
        \textbf{Category} & \textbf{Description}  & \textbf{Example} & \textbf{Frequency}\\
        \midrule
        User Question & Queries or questions provided by users. & \{\{question\}\}, \{\{query\}\} & 24.5\%\\
        \rowcolor{gray!10} Contextual Information & Background or supplementary input that helps set the stage for the task but is not the primary focus. & \{\{chat\_history\}\}, \{\{background\_info\}\} & 19.5\%\\
        Knowledge Input  & The core content that the prompt directly processes or manipulates. & \{\{document\}\}& 50.9\%\\
        \rowcolor{gray!10} Metadata/Short Phrases & Brief inputs or settings that define specific parameters or goals for the task. & \{\{language\}\}, \quad \{\{username\}\} & 43.4\% \\
        \bottomrule
    \end{tabular}
}
\end{table}

\subsubsection{Positional Distribution of Placeholder}\ 

\begin{figure}[t]
    \centering
  \includegraphics[width=\linewidth]{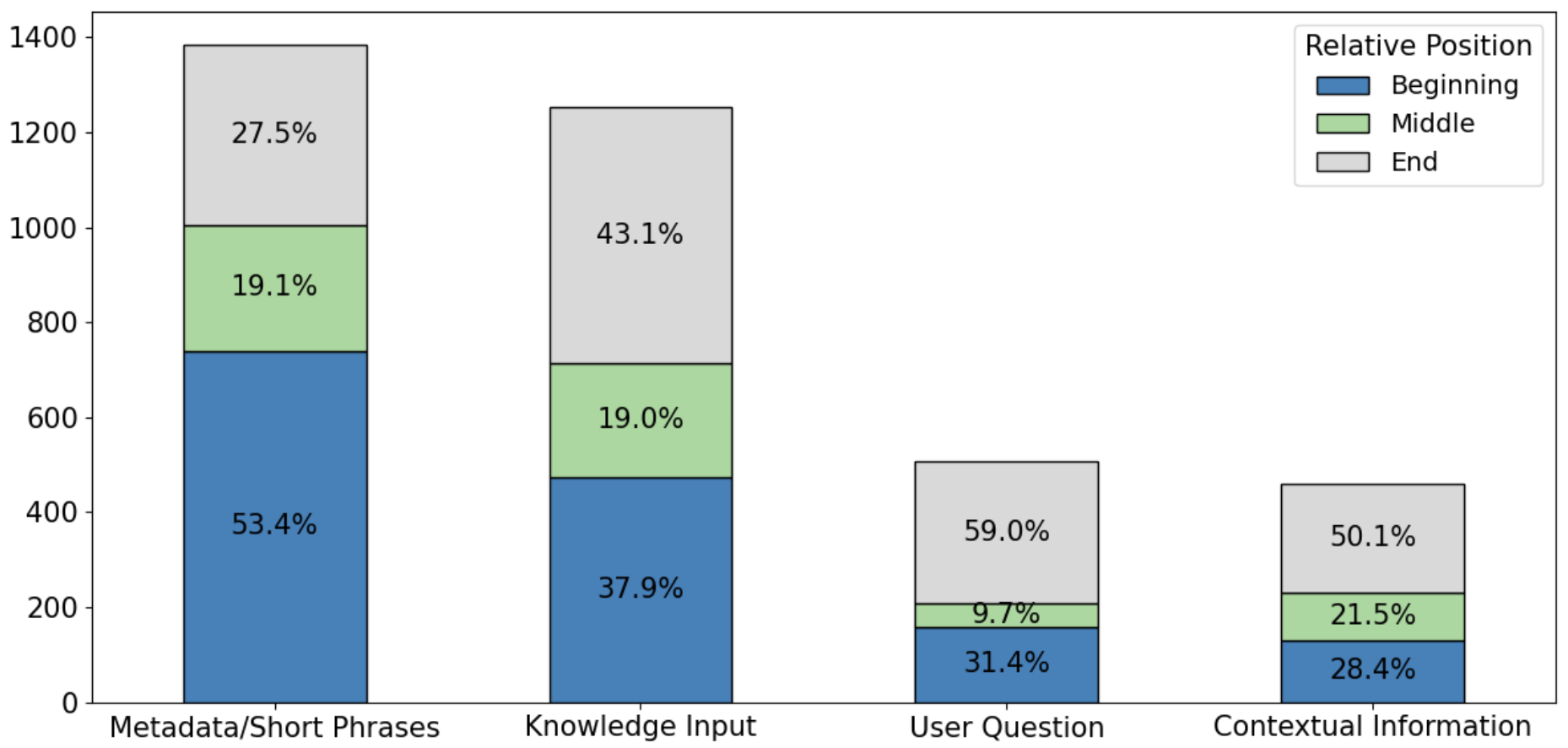}
    \caption{Frequency distribution of placeholder positional occurrences for different placeholder types.}
    \vspace{-3mm}
    \label{fig:placeholder_position}
\end{figure}

\begin{figure}[t]
    \centering
  \includegraphics[width=\linewidth]{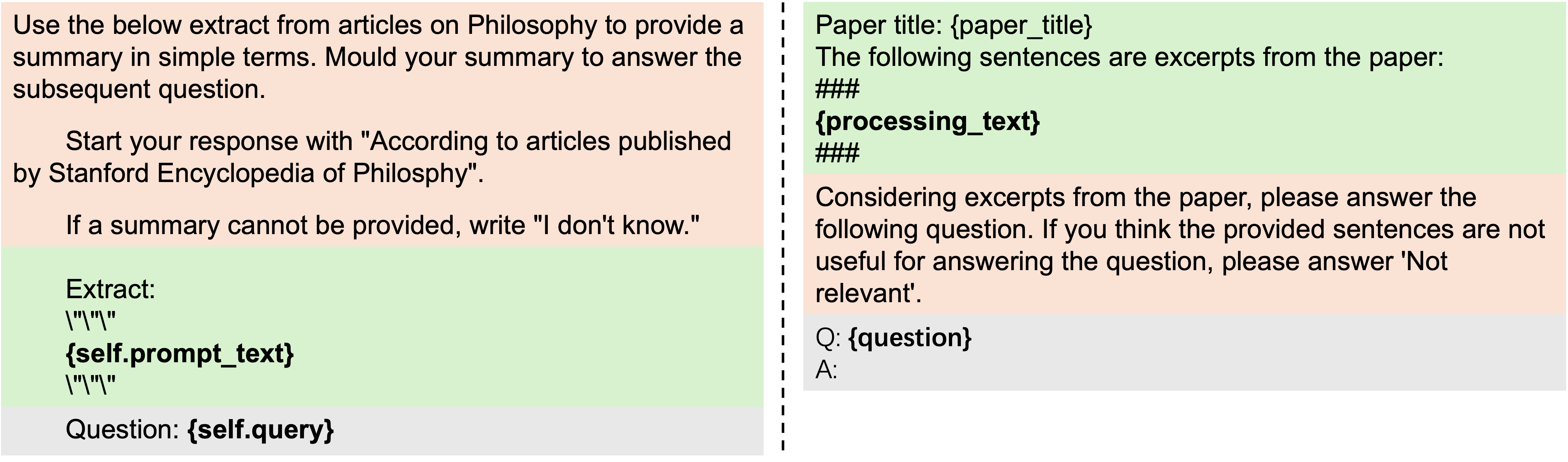}
    \caption{Examples of prompt templates with different Knowledge Input positions~\cite{thinkai, AI-LaBuddy}.}
    \vspace{-3mm}
    \label{fig:placeholder_example}
\end{figure}

\noindent We analyze the positional distribution of placeholders within prompt templates, dividing each template into three sections—beginning, middle, and end—each representing one-third of the template's length. Figure~\ref{fig:placeholder_position} illustrates this distribution, in which approximately 60\% of user questions appear at the end, reflecting a common structural pattern, whereas \{Knowledge Input\} is more evenly distributed between the beginning and end positions. Figure~\ref{fig:placeholder_example} presents examples with varied \{Knowledge Input\} positions. Additionally, placeholder content length varies considerably; longer knowledge inputs may lead to information loss in LLMs across extended prompts.

\begin{tcolorbox}[colback=black!5!white,colframe=black!75!black,before upper=\vspace{-.25cm},after upper=\vspace{-.25cm},left=0.1pt, right=0.1pt]
  \textbf{Finding 5: } There are four types of placeholder, and ``Knowledge Input'' is the most frequently used in prompt templates, and developers always place it either at the beginning or the end of templates. 
\end{tcolorbox}

\subsubsection{Placeholder Name}\label{section:placeholder_name}\

\noindent One key aspect of placeholder design is naming, which plays a similar role as the variable name in code. Clear and descriptive names are essential to understand the intent of placeholders.
Notably, placeholder names like ``text'' (4.44\%) and ``input'' (2.35\%) are often used. Among these, ``text'' is the second most frequently used placeholder name behind ``question'' (6.18\%), highlighting a broader issue in placeholder naming. These names are general and do not precisely indicate their specific context, leading to potential confusion for developers or users. For example, ``text'' could refer to any kind of input text, so developers and users cannot directly understand what they need to input through its name. In this case, more descriptive placeholder names are suggested so that developers and users can easily understand what they need to input without considering the whole prompt template context.

\begin{tcolorbox}[colback=black!5!white,colframe=black!75!black,before upper=\vspace{-.25cm},after upper=\vspace{-.25cm},left=0.1pt, right=0.1pt]
  \textbf{Finding 6: } Similar to the variable naming convention, non-semantic placeholder names such as ``text'' and ``input'' are still commonly used in prompt templates, hindering prompt understanding and maintenance.
\end{tcolorbox}

\subsection{RQ3: Evaluating Patterns through Sample Testing}\label{section:RQ3}




\subsubsection{Json Output Format}\ 

To investigate the effect of different JSON output patterns, we test three identified patterns displayed in Figure~\ref{fig:json_format}. Five representative templates targeting different tasks (\eg, tweet analysis, DNS parsing) are selected for each pattern and three diverse input instances are generated per template, yielding a total of 45 templates after reformatting them to fit all patterns. We define two metrics rated from 1 to 5 to evaluate the JSON object in LLM output~\cite{xia2024fofo, Joshi2024CoPrompterUE}:
\begin{itemize}[leftmargin=0.7cm]
\item \textit{Format Following:} Measures the consistency of generated JSON strings with the defined format, including attribute count, names, and structural uniformity across outputs. 

\item \textit{Content Following:} Assesses the alignment of generated content with the intent in the prompt template.
\end{itemize}

\begin{table}[t]
\caption{\rmfamily LLM output quality under three json output format patterns.}
\rmfamily
\centering
\tiny
\resizebox{\linewidth}{!}{
\label{tab:json_format}
\begin{tabular}{lcccc}
\toprule
   \multirow{2}{*}{\textbf{Pattern}}  & \multicolumn{2}{c}{\textbf{Format Following}}& \multicolumn{2}{c}{\textbf{Content Following}}\\
    & \llamaversion & \gptversion & \llamaversion & \gptversion \\
\midrule
Pattern 1 & 3.09  & 3.21 & 3.70  & 3.50  \\
Pattern 2 & 4.66  &  4.86  & 4.02   & 4.30   \\
Pattern 3 & \textbf{4.90}   & \textbf{4.96}  & \textbf{4.47} & \textbf{4.53}\\
\bottomrule
\end{tabular}
}
\begin{tablenotes}
\footnotesize
\item Pattern 1 is Json Output, Pattern 2 is Json Output + Json Attribute Name,
Pattern 3 is Json Output + Json Attribute Name + Json Attribute Description.
\end{tablenotes}
\end{table}

As shown in Table~\ref{tab:json_format}, both models exhibit similar trends, with Pattern 3 achieving the highest scores across metrics and models. In Format-Following, Patterns 2 and 3 outperform Pattern 1, which scores lower due to inconsistencies in attribute count and naming, stemming from its lack of explicit definitions. Patterns 2 and 3 mitigate these issues through explicit ``Json Attribute Names'', ensuring greater structural consistency.

For Content-Following, Pattern 3 achieves the highest scores, with \llamaversion\ at 4.47 (+0.45) and \gptversion\ at 4.53 (+0.23), suggesting that detailed ``Json Attribute Descriptions'' enhance the model's ability to generate content that aligns closely with user requirements. These findings underscore the value of attribute-level detail in improving both precision and semantic adherence.



\begin{figure}[t]
    \centering
     \includegraphics[width=\linewidth]{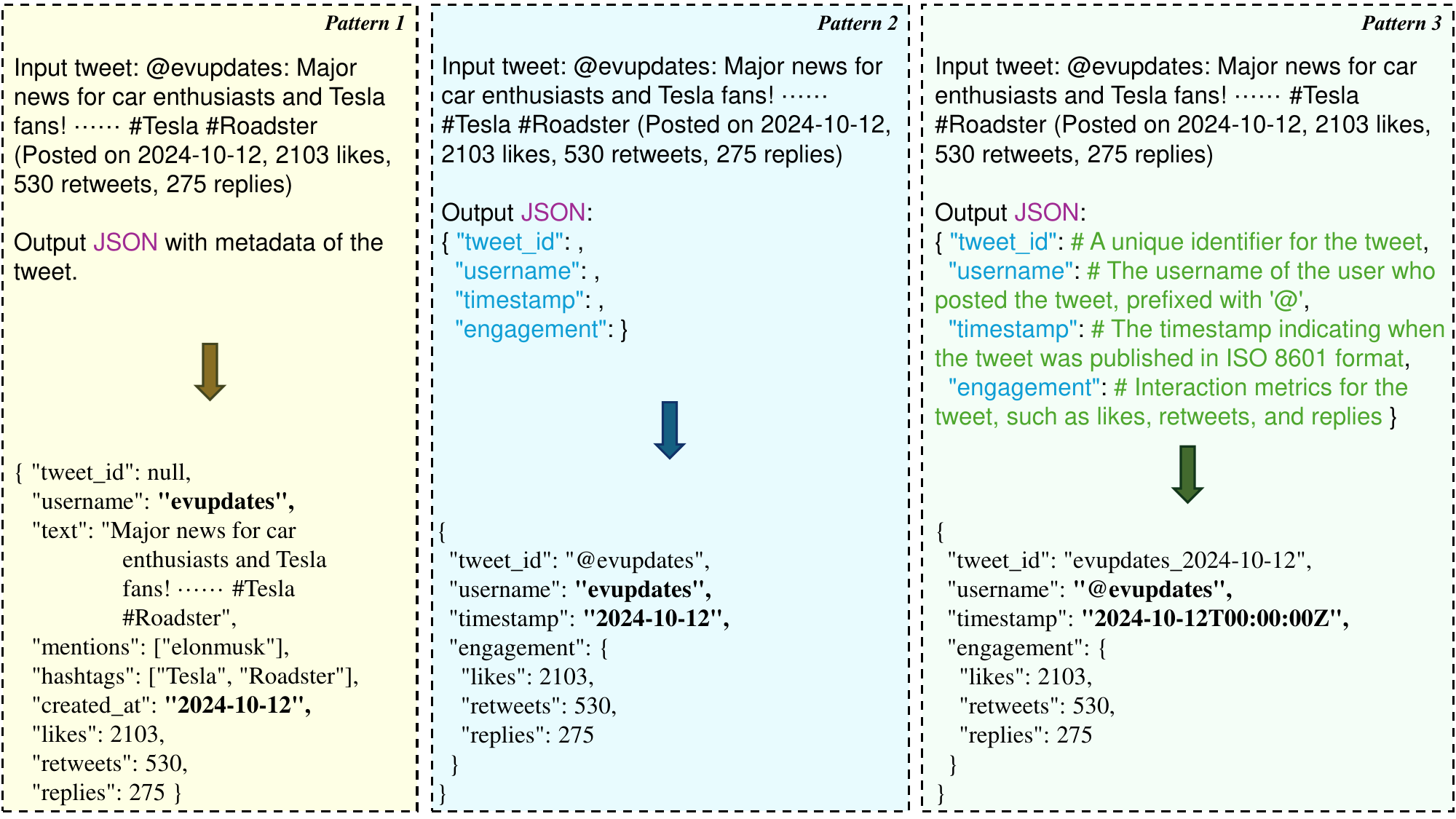}
    \caption{Output examples of different json output format patterns~\cite{dev-gpt}.}
    \vspace{-3mm}
    \label{fig:json_example}
\end{figure}

Figure~\ref{fig:json_example} illustrates an example of extracting tweet information using three prompt templates with identical input text but differing JSON output formats. Case 1 broadly requests a JSON output without defining required fields, leading to redundant outputs like ``text'', ``mentions'', and ``hashtags''. Case 2 specifies field names like ``tweet\_id'' and ``engagement'', producing more structured outputs, but attributes like ``username'' and ``timestamp'' remain ambiguous without further clarification. Case 3 resolves these ambiguities by adding clear descriptions for each attribute (e.g., defining ``username'' as the Twitter handle prefixed with ``@'' and specifying ``timestamp'' to follow the ISO 8601 format~\cite{ISO_8601}), resulting in outputs that are more accurate and aligned with user expectations.

\begin{tcolorbox}[colback=black!5!white,colframe=black!75!black,before upper=\vspace{-.25cm},after upper=\vspace{-.25cm},left=0.1pt, right=0.1pt]
  \textbf{Finding 7: } When generating LLM outputs in JSON format, explicit Json Attribute Names enhance format consistency by ensuring structured and uniform outputs. Additionally, detailed Json Attribute Descriptions further refine content-following, reducing ambiguity and better aligning with user-defined requirements.
\end{tcolorbox}

\subsubsection{Exclusion Constraint for Output}\ 

Well-defined JSON output formats enhance the format-following abilities of LLMs when generating JSON objects. However, as observed in the previous experiment, LLMs often include redundant explanations alongside the JSON object, leading to output inconsistency. This experiment evaluates the effectiveness of exclusion constraints in reducing such redundancies. Using the same 15 populated templates from the JSON Output Pattern 3 experiment, we apply an exclusion constraint, ``Do not provide any other output text beyond the JSON string'', positioned before the JSON format definition as per the component order identified in findings 1. For evaluation, we define the metric \textit{Format Following} as a binary value: ``1'' if the output consists solely of the JSON string and is ready for direct parsing, and ``0'' otherwise.





Results highlight the impact of the exclusion constraint. For the \llamaversion\ model, the original prompts yield a Format Following rate of 40\% (only JSON string in 40\% of outputs). The constraint raises this rate to 100\%, demonstrating improved adherence. 
In 16.67\% of cases, the output is enclosed in """json""" without any other explanation text, which we do not consider redundant.
The \gptversion\ model performs better with the original prompts, achieving an 86.67\% adherence rate, further increasing to 100\% with the exclusion constraint applied. These findings underline the exclusion constraint's value in improving clarity, reducing redundancy, and ensuring strict adherence to output format requirements.

\begin{tcolorbox}[colback=black!5!white,colframe=black!75!black,before upper=\vspace{-.25cm},after upper=\vspace{-.25cm},left=0.1pt, right=0.1pt]
  \textbf{Finding 8: } Using only JSON format definitions is insufficient to fully prevent extraneous explanations or comments. Combining ``Do'' instructions, such as explicit output format definitions, with ``Don't'' instructions, like exclusion constraints, significantly reduces redundancy while maintaining high output consistency in LLM-generated content.
\end{tcolorbox}

\subsubsection{Position of Knowledge Input Placeholder}\label{section:Position of Knowledge Input Placeholder}\ 

In this experiment, we explore the optimal position of the \{Knowledge Input\} placeholder, the most commonly used type, as identified in Finding 5. We use a retrieval-augmented generation (RAG)-style task where users pose questions, external knowledge is provided as input, and developers define instruction (\eg, directive, constraint, output format) to process the dynamic input. To assess the impact of the \{Knowledge Input\} placeholder’s position, placed either at the beginning or end of the prompt, as noted in Finding 5, we test two configurations: Instruction First, where the task-intent portion of the instruction precedes the knowledge input, and Placeholder First, where the knowledge input precedes the instruction. In both configurations, the \{User Question\} placeholder remains at the end, following the most frequent position of \{User Question\} in Figure~\ref{fig:placeholder_position}. The populated templates span a variety of topics, such as medical and code-related QA. We also select inputs of varying lengths to adapt to real-world usage scenarios. Evaluation is based on two human-assessed metrics, scored on a 1–5 scale, which assesses the relevance to both the developer-defined task instruction and the end-user question in general task-oriented Q\&A systems~\cite{luo2022critical}:
\begin{itemize}[leftmargin=0.7cm]
\item \textit{Content Following (Question): } Alignment with user question.

\item \textit{Content Following (Task Intent): } Adherence to the developer-defined task intent (\eg, including directive, constraint, and output format).
\end{itemize} 

Table~\ref{tab:placeholer_avg_length} presents the average Content Following scores under different configurations. Both models exhibit high scores for Content Following (Question), reflecting consistent alignment with user queries at the end of the prompt template regardless of placeholder position. However, for Content Following (Task Intent), the Placeholder First pattern consistently outperforms the Instruction First pattern, with average scores of 4.63 for LLaMA (+0.91) and 4.60 for GPT (+0.34). Notably, as knowledge input length increases, the Instruction First pattern suffers a more significant performance drop compared to the Placeholder First pattern. This suggests that placing the task intent of instruction before the knowledge input increases the likelihood of forgetting or misalignment as input length grows. Conversely, placing the task intent of instruction after the knowledge input mitigates this issue, maintaining robust task intent adherence even with long inputs.

\begin{table}[htpb]
\caption{\rmfamily Output quality under different position patterns}
\vspace{-3mm}
\rmfamily
\centering
\tiny
\resizebox{\linewidth}{!}{
\label{tab:placeholer_avg_length}
\begin{tabular}{llcc}
\toprule
\multirow{2}{*}{\textbf{Pattern}}        & \multirow{2}{*}{\textbf{Model}} & \textbf{Content Following} & \textbf{Content Following} \\ 
        &  & \textbf{(Question)} & \textbf{(Task Intent)} \\ 
\midrule
\multirow{2}{*}{Instruction First} & LLaMA        & 4.52                                  & 3.72                                      \\
                           & GPT          & 4.61                                  & 4.26                                      \\ 
\midrule
\multirow{2}{*}{Placeholder First} & LLaMA        & \textbf{4.84} (+0.32)                         & \textbf{4.63} (+0.91)                            \\
                           & GPT          & \textbf{4.70} (+0.09)                        & \textbf{4.60} (+0.34)                            \\ 
\bottomrule
\end{tabular}
}
\begin{tablenotes}
\footnotesize
\item Instruction First Pattern is Task Intent -> \{Knowledge Input\} -> \{User Question\}. Placeholder First Pattern is \{Knowledge Input\} -> Task Intent -> \{User Question\}.
\end{tablenotes}
\end{table}

\begin{figure}[htpb]
    \centering
  \includegraphics[width=\linewidth]{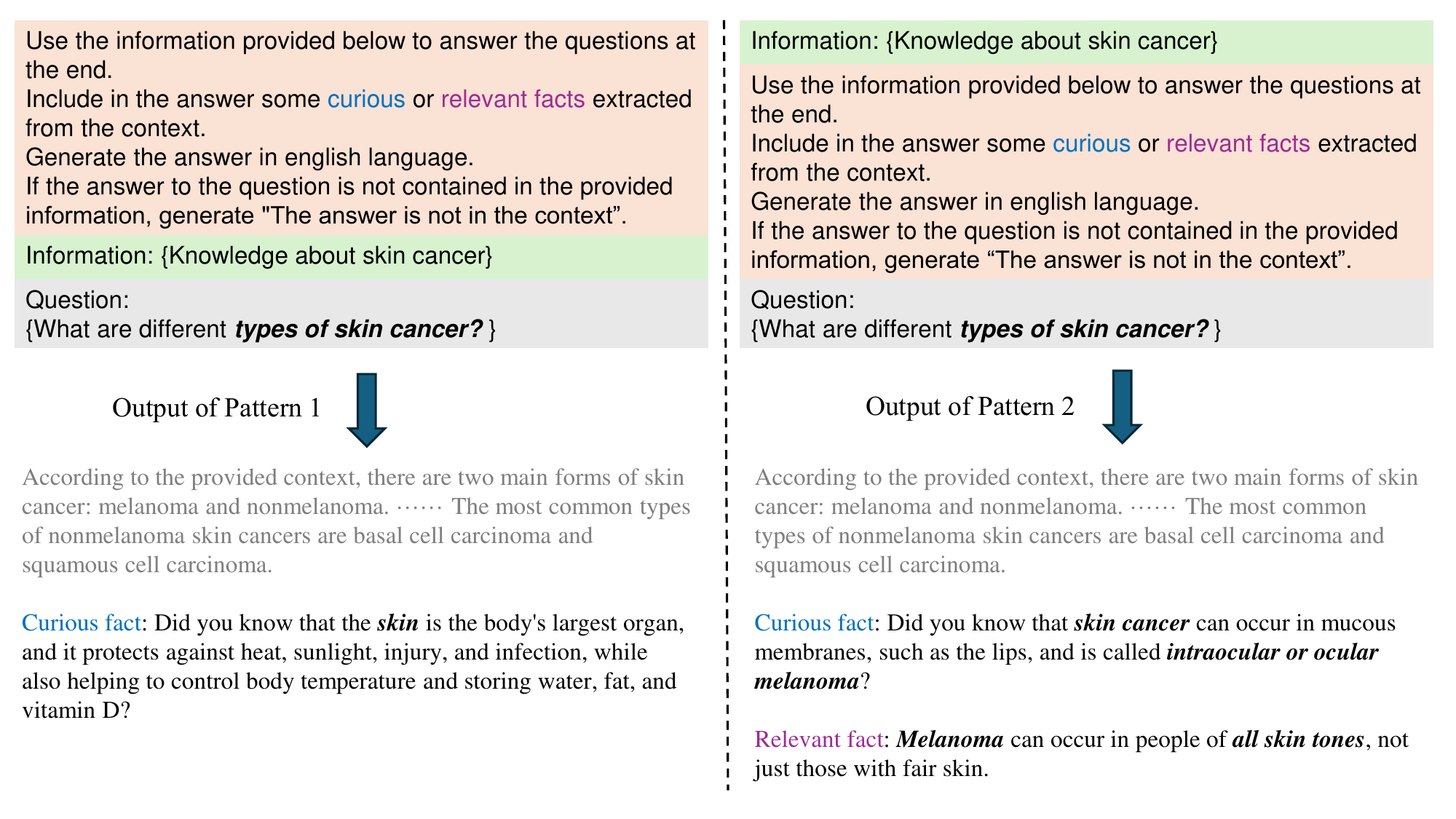}
    \caption{Output examples with {Knowledge Input} in different prompt template positions~\cite{wikisearch}. }
    \vspace{-5mm}
    \label{fig:placeholder_length}
\end{figure}

Figure~\ref{fig:placeholder_length} presents a comparison using a long input text. The left example follows the Instruction First pattern, while the right one follows Placeholder First. In both, the user question remains at the end of the prompt template. Colored sections highlight the instruction (red), knowledge input (green), and user question (grey). The upper portion of each side presents the filled prompt template, while the lower portion showcases the respective LLM output. In this example, the long input is a detailed description of skin cancer. Regarding the LLM outputs, both patterns provide similar responses to the user question, with the middle portion of the long description about skin cancer being reduced in both cases. However, the differences arise in how the LLM addresses the facts required by the instruction. In the Instruction First pattern, the ``curious fact'' output discusses what the skin is, which, although interesting, is unrelated to the user question. Additionally, the relevant fact is missing. In contrast, the Placeholder First pattern generates more aligned results. The ``curious fact'' and ``relevant fact'' both relate to skin cancer types, discussing where they occur (\eg, on mucous membranes) and their occurrence across various skin tones, thus providing a more relevant and comprehensive response to the instruction and user question.
\vspace{-1mm}

\begin{tcolorbox}[colback=black!5!white,colframe=black!75!black,before upper=\vspace{-.25cm},after upper=\vspace{-.25cm},left=0.1pt, right=0.1pt]
  \textbf{Finding 9: } Positioning the task intent of instruction and user question after the knowledge input enhances output consistency and mitigates information decay in Q\&A tasks, particularly when processing prompt templates with highly variable input lengths in \llmapps.
\end{tcolorbox}
\vspace{-3mm}


\section{Implications}
Our research offers actionable insights into prompt engineering for various parties in the software industry:
\subsection{Implications for LLM Providers}
LLM providers can enhance the usability and performance of their APIs by offering best practices for designing, testing, and optimizing prompt templates. 
Almost all commercial LLM providers support prompt templates, such as Google Cloud Gemini~\cite{gemini_prompt_templates} and Langchain~\cite{langchain_prompt_templates}. However, none of them provides any guideline on how to write effective prompt templates. 

\textbf{Pre-defined prompt templates.} To address the challenges developers face in crafting effective prompts, LLM providers could offer pre-defined templates for common tasks. For instance, templates could include an information-seeking task with JSON as the output format (see Section~\ref{section:component_content}) or a RAG-style question-answering task (see Section~\ref{section:Position of Knowledge Input Placeholder}). By leveraging the overall component order identified in Section~\ref{section:Component Order}, providers can establish a unified structure for these templates. Furthermore, based on the pattern testing results outlined in Section~\ref{section:RQ3}, API providers can adopt optimal writing patterns tailored to specific tasks, ensuring improved generation performance for these pre-defined templates.


\textbf{Automated template evaluation/explainability tools.} Informed by our findings in RQ3, API providers could develop automated evaluation tools to assist \llmapp\ developers in testing and refining prompt templates. These tools should enable developers to compare outputs from different template patterns for the same inputs, facilitating the identification of optimal designs. Moreover, the tools should analyze the template structure to identify measurable evaluation criteria, such as constraints or output requirements. These evaluations could give scores and explanations based on metrics like content-following and format-following. Additional features, such as model version comparisons and prompt template history tracking, could further enhance the usability of these tools.

\vspace{-3mm}

\subsection{Implications for \llmapp\ Developers}
After an \llmapp's release, maintaining prompt templates based on user feedback is crucial to ensure high-quality outputs. However, identifying the most effective improvement strategy—such as modifying specific components or adopting new prompt techniques—can be challenging. While switching to a more powerful model may improve performance, it also significantly increases costs, making cost efficiency a critical consideration for the success of \llmapps.

\textbf{Prompt templates maintenance. } Prompt templates should adapt dynamically to enhance user experience, incorporating user feedback and expert reviews for continuous refinement~\cite{Macneil2023PromptMM, Joshi2024CoPrompterUE}. Analyzing historical usage data, such as input lengths and content types, helps developers optimize placeholders and adjust component positions to prevent key information from being overlooked, especially when handling long inputs (as noted in RQ3). Placeholders should also align with real-world scenarios—for instance, separating background and analytical inputs into distinct placeholders improves clarity and usability. Incorporating metadata placeholders, such as \{output\_format\}, ensures flexibility and robustness, accommodating diverse user needs. Developer and expert reviews further validate refinements, aligning templates with best practices. Additionally, as observed in Section~\ref{section:placeholder_name}, many prompt templates still use ambiguous placeholders (\eg, approximately 5\% of placeholders are named simply as ``text''), which lacks meaningful context and can complicate maintenance during \llmapp\ evolution. Clear, descriptive naming reduces errors, mitigates challenges arising from memory limitations and developer turnover, and ensures long-term software reliability.

\textbf{Using well-defined prompt templates to strengthen weak LLMs. } As demonstrated in our sample testing results in Section~\ref{section:RQ3}, well-defined prompt templates significantly enhance the instruction-following capability of weaker models (\eg, \llamaversion). In some cases, these templates enable weaker models to achieve performance levels comparable to the best-performing LLMs (\eg, \gptversion). For instance, in the long-input experiment discussed in Section~\ref{section:Position of Knowledge Input Placeholder}, the output quality boost achieved with a well-defined prompt template for \llamaversion\ was nearly double that of \gptversion. This highlights the critical role of carefully designed prompt templates for optimizing weaker models. When selecting foundation models for \llmapps, developers should first consider re-designing prompt templates for the target task if the generation quality falls short of expectations, rather than immediately switching to a more advanced model. Well-designed prompt templates can significantly strengthen the instruction-following abilities of weaker models, therefore contributing to reducing costs, an essential factor for business success.

\textbf{Trade-offs in In-Context learning. } In-context learning has become a widely adopted prompt engineering technique in software engineering research~\cite{hou2024large}. However, our statistical analysis of component distribution in prompt templates (Table~\ref{tab:component_distribution}) reveals that fewer than 20\% of applications in our dataset incorporate few-shot examples in their prompts. Considering the usage of dynamically loaded examples through placeholders (less than 5\%), few-shot examples are still not commonly used in prompt templates. Prior research also highlights that clearly defined tasks without examples can sometimes outperform those with few-shot examples in terms of generation quality~\cite{reynolds2021prompt}. This finding likely explains why many \llmapps\ omit few-shot learning. While few-shot examples can sometimes enhance performance, they are often unnecessary when a well-defined prompt template is used. Moreover, including such examples can introduce drawbacks, such as increased token costs and the risk of semantic contamination. In-context learning is not a one-size-fits-all solution. When designing \llmapps, developers should prioritize refining and optimizing prompt templates to achieve clarity and alignment with task requirements rather than defaulting to few-shot learning.

\vspace{-1mm}

\section{Related Work}



\noindent \textbf{Software Engineering for LLM.} The emerging field of Software Engineering for Large Language Models (SE4LLM) applies software engineering principles to various stages of LLM development, addressing challenges such as efficiency through system stack optimization~\cite{aminabadi2022deepspeed}, model compression via prompt learning~\cite{xu2023compress}, and security with multi-round automatic red-teaming~\cite{ge2023mart} and monitoring through modular approaches~\cite{goyal2024llmguard}. Research has also explored interpretability using attention visualization~\cite{galassi2020attention}, concept-based analysis~\cite{oikarinen2023label}, and subnetwork extraction~\cite{wang2020interpret}.

Beyond LLM optimization, attention has turned to \llmapps, focusing on post-deployment challenges. Zhao \etal~\cite{zhao2024llm} examined LLM app stores, investigating user experience, developer strategies, and ecosystem dynamics while highlighting challenges such as security and privacy. Security issues in the \llmapp\ ecosystem, including jailbreaking, prompt injection, credential leaks, and inconsistent data provision, have been extensively studied~\cite{Deng2023MASTERKEYAJ, liu2023prompt, yan2024exploring}, emphasizing the need for robust safeguards.

In contrast to these post-deployment studies, our research focuses on the pre-deployment phase of \llmapp\ development by analyzing prompt templates—the critical interface between users and LLMs. By examining prompt structure, components, and patterns, we provide actionable insights to help developers design more effective templates, thereby optimizing interactions before model execution. This work offers foundational support for enhancing both the usability and performance of \llmapps.

\noindent \textbf{Prompt Engineering. } Prompt engineering is critical for guiding LLMs, but the variability of natural language poses challenges in achieving clarity and consistency. Recent research has focused on common elements and structural patterns, including standardizing prompt elements into reusable frameworks~\cite{Wang2024LangGPTRS}, emphasizing task-specific instructions and formatting~\cite{reynolds2021prompt}, and analyzing the effects of component sequencing~\cite{liu2023instruction, cuconasu2024power, zhao2021calibrate}. Other studies have explored the impact of minor prompt modifications, such as punctuation~\cite{sclar2023quantifying} and word rephrasing~\cite{yang2024just}.

In contrast to studies on general prompt design, our research focuses on systematically analyzing prompt templates used in \llmapps. Building on prior work in prompt analysis~\cite{schulhoff2024prompt, GoogleCloud}, we refine the analysis framework by targeting the common components of prompt templates. Unlike general prompts crafted for direct user-LLM interactions, prompt templates designed by \llmapp\ developers are typically longer and more intricate, often incorporating multiple components to articulate detailed task instructions~\cite{wang2022super}. A distinctive feature of prompt templates is the inclusion of placeholders~\cite{schulhoff2024prompt}, which general prompts usually lack. Placeholders allow developers to account for diverse user inputs and scenarios while enhancing the reusability and adaptability of templates. To the best of our knowledge, this is the first study to systematically analyze prompt templates in \llmapps, providing developers with actionable guidance for designing effective, reusable templates that accommodate a broad range of user inputs.

\section{Conclusion}
This paper provides a comprehensive analysis of prompt template structure and composition in LLM-powered applications, analyzing how developers design these templates to optimize LLM's instruction-following abilities. We construct a dataset of prompt templates from GitHub open-source projects, identifying common components and placeholders in those prompt templates. Through LLM-assisted and human-verified analysis, we analyze the frequently used terms of these components and placeholders, as well as their positions, and further identify several organizational patterns based on the analysis result. Finally, we conduct prompt template testing to evaluate how different patterns influence the instruction-following performance of LLM. These findings offer foundational insights for prompt engineering, guiding the design of robust prompt templates that enhance the quality of LLM outputs across various \llmapps.

\section*{Acknowledgements}

This research was supported by OpenAI through the provision of API credits under OpenAI Researcher Access Program. We appreciate their support in facilitating our study on prompt template analysis.

\newpage

\bibliographystyle{ACM-Reference-Format}
\bibliography{ref}

\end{document}